# L'engagement logistique des acteurs des chaînes courtes alimentaires comme creuset d'alternativité ?

Camille Horvath
Auteur correspondant
Chaire Logistics City/LVMT– Université Gustave Eiffel
Camille.horvath@inrae.fr
0000-0002-8687-4176

Céline Raimbert
AME/SPLOTT – Université Gustave Eiffel
celine.raimbert@inrae.fr
0009-0003-8588-7068

Gwenaëlle Raton
Université Gustave Eiffel, SPLOTT
gwenaelle.raton@univ-eiffel.fr
0000-0002-9779-2192

**Résumé :**

Face aux externalités négatives (environnementales, sociales et économiques) du modèle alimentaire dominant, un consensus émerge sur la nécessité de développer des systèmes alimentaires alternatifs capables de favoriser une transition vers un modèle plus durable. Parmi ces alternatives, les circuits courts alimentaires (CC) sont identifiés comme des creusets potentiels d'alternativité. Définis comme des circuits de distribution comprenant un intermédiaire maximum entre le producteur et le consommateur, les CC peuvent contribuer à la reterritorialisation des flux alimentaires et à la création de proximités (relationnelles, géographiques) entre les acteurs de la chaîne alimentaire. Cette étude examine comment se traduit l'engagement logistique dans les choix d'un point de vente en CC des consommateurs et des agriculteurs. Elle évalue cet engagement à travers les temps de trajet que les agriculteurs et les consommateurs sont prêts à parcourir pour vendre ou acheter dans un point de vente en CC, ainsi que le prix auquel ils sont prêts à vendre ou acheter des produits en CC. Une expérience de choix discrets menée en 2022 auprès de 154 chefs d'exploitation agricole et 1022 consommateurs a permis d'analyser les préférences en matière de types de points de vente, de prix de vente, de temps de trajet et de proximité relationnelle. Les résultats montrent que les consommateurs engagés dans les CC privilégient les points de vente alternatifs tels que les fermes et les associations, même pour des temps de trajet plus longs. Les agriculteurs valorisent également ces points de vente, bien que l'engagement logistique soit plus marqué chez ceux qui commercialisent déjà en CC. Le supermarché apparaît comme une option moins attrayante pour les profils alternatifs mais constitue une porte d'entrée pour les nouveaux participants aux CC. La proximité relationnelle entre agriculteurs et consommateurs est essentielle, avec une préférence pour les échanges directs lors de la vente. Cette étude souligne l'importance de l'engagement des acteurs dans les CC et offre des perspectives pour le développement d'une logistique alternative, en valorisant la proximité relationnelle et en utilisant des points de vente conventionnels comme porte d'entrée pour de nouveaux acteurs dans les CC.

**Abstract :**

In response to the negative environmental, social, and economic externalities of the dominant food model, a consensus is emerging on the need to develop alternative food systems capable of facilitating a transition to a more sustainable model. Among these alternatives, short food supply chains (SFSCs) are identified as potential crucibles of alternativity. Defined as distribution channels with a maximum of one intermediary between producer and consumer, SFSCs can contribute to the reterritorialization of food flows and the creation of relational and geographical proximities among food chain actors. This study examines how logistical engagement translates into the choice of SFSC points of sale by consumers and farmers. It evaluates this engagement through the travel times farmers and consumers are willing to endure to sell or buy at an SFSC point of sale, and the prices they are willing to accept or pay for SFSC products. A discrete choice experiment conducted in 2022 with 154 farm managers and 1022 consumers analyzed preferences regarding types of points of sale, sales prices, travel times, and relational proximity. The results show that consumers engaged in SFSCs favour alternative points of sale such as farms and associations, even for longer travel times. Farmers also value these points of sale, with logistical engagement more pronounced among those already marketing through SFSCs. Supermarkets appear less attractive to alternative profiles but serve as entry points for new participants in SFSCs. Relational proximity between farmers and consumers is important, with a preference for direct exchanges during sales. This study highlights the importance of actor engagement in SFSCs and offers perspectives for developing alternative logistics by valuing relational proximity and using conventional points of sale as entry points for new actors in the SFSCs.

## 1. Introduction

Au vu des externalités négatives (environnementales, sociales et économiques) produites par le modèle agri-alimentaire dominant, l'émergence de systèmes alternatifs, capables de susciter une transition alimentaire pourrait permettre de jeter les bases d'un modèle plus durable (Berti 2020 ; IPES-Food 2018). Parmi les actions en ce sens, les circuits courts alimentaires[1] (CC) sont identifiés comme un potentiel creuset d'alternativité, du fait de leur capacité à favoriser une reterritorialisation des flux alimentaires, mais aussi le développement de proximités entre acteurs de l'alimentation, et notamment entre producteurs agricoles et consommateurs (Bui et al., 2016; Chiffoleau and Dourian, 2020; Kirwan et al., 2013; Seyfang, 2006). Ils portent, en cela, ce que Le Velly (2017) nomme une « *promesse de différence* ». Pour les consommateurs, la différence promise par les CC concerne, entre autres, une perception d'une meilleure qualité et moindre standardisation des produits (Burchardi et al., 2005; Meyerding et al., 2019; Thilmany et al., 2008), ainsi qu'une reconnexion avec les producteurs

---
[1] Définis par le plan Barnier (2009) comme des circuits de commercialisation avec un intermédiaire au maximum

agricoles et un soutien à l'économie locale (Burchardi et al., 2005; Meyerding et al., 2019; Pernin and Petitprêtre, 2013; Skallerud and Wien, 2019). Elle s'applique aussi aux producteurs agricoles, pour lesquels les CC peuvent représenter une opportunité de meilleure rémunération, davantage d'autonomie, ou encore une plus grande proximité avec les consommateurs et leurs attentes (Alonso Ugaglia et al., 2020; Samak, 2012).

Pour autant, vendre ou acheter en CC impose des contraintes liées, entre autres, au temps nécessaire pour se rendre dans un point de vente en CC. Pour les consommateurs, la littérature montre que l'accessibilité aux points de vente est déterminante, puisque les consommateurs des CC consomment avant tout « dans leur quartier » (Prigent-Simonin et al., 2012). Sachant, par ailleurs, que ces consommateurs continuent d'acheter en circuits longs (CRIOC, 2010 ; CREDOC, 2017), on peut considérer que l'achat en CC suppose un trajet supplémentaire, en plus de celui pour acheter en Grandes et Moyennes Surfaces (GMS), fréquentées par plus de 90 % des consommateurs français. Pour les agriculteurs, les travaux sur les enjeux logistiques des CC montrent que ce sont majoritairement eux qui se chargent de la livraison de leurs produits. Cette activité peut se révéler d'autant plus chronophage pour l'exploitation que les agriculteurs ne sont pas, *a priori*, des professionnels de la logistique (Blanquart et al., 2015; Raton et al., 2015). Livrer en CC implique ainsi d'importants coûts logistiques (y compris la valeur du temps) qui peuvent avoir un poids sur la rentabilité des exploitations en CC (Raton et al., 2018). Par ailleurs, Raton et Raimbert (2019) ont montré que les décisions stratégiques des agriculteurs en CC concernant la définition de leur zone de chalandise et la distribution spatiale de leurs débouchés prennent en compte les contraintes logistiques, spécifiques au tissu commercial local[2].

On comprend, dès lors, que recourir aux CC, et participer au développement de systèmes agri-alimentaires alternatifs, exige des changements de pratiques et d'organisation, qui ne relèvent pas seulement de pratiques de production et de consommation, mais aussi des pratiques de mobilité (i.e des trajets nécessaires pour se rendre dans un point de vente en CC). Nous formulons l'hypothèse que les temps dédiés aux livraisons et à l'approvisionnement sont centraux dans les choix réalisés par les agriculteurs, et que leur consentement à y consacrer plus ou moins de temps repose sur leur engagement en faveur de pratiques alternatives. De la même manière, pour les consommateurs, le fait de prendre plus ou moins temps pour aller acheter des produits issus des CC peut indiquer leur engagement dans ces systèmes alternatifs. Le cas de certains groupements d'achat fournit un exemple abouti d'engagement : les consommateurs vont eux-mêmes récupérer des produits dans les fermes, qu'ils rendent ensuite disponibles à d'autres consommateurs. La répartition plus juste des tâches, et donc des coûts et des temps de trajets acceptés par les consommateurs et les agriculteurs, peut être vu comme un indicateur de l'engagement, qui se définirait alors comme un engagement logistique. La question qui se pose alors est la suivante : dans quelle mesure le degré d'alternativité des points de vente en CC influe sur l'engagement logistique des producteurs agricoles et des consommateurs ? L'originalité de ce travail est qu'il traite de l'alternativité des systèmes agri-alimentaires, non pas du point de vue des changements des pratiques de production des uns ou de consommation des autres, mais plutôt de leurs pratiques logistiques (trajets pour se rendre dans un point de vente en CC).

---

[2]Cette étude de cas dans les Hauts-de-France met en lumière la diversité des stratégies des agriculteurs en CC : choisir de livrer en ville, parfois beaucoup plus loin, pour concentrer les débouchés dans un bassin de consommation dense, ou, au contraire, pour éviter la congestion, privilégier les villes moyennes ou petites. Dans ce cas, les agriculteurs préfèrent des bassins de consommation moins denses/plus fragmentés, en multipliant les débouchés et/ou en étendant leur zone de livraison.

Pour saisir plus finement les leviers de cet engagement, nous mettons en regard le poids de la proximité spatiale (autrement dit le temps de trajet, associé à une distance) avec d'autres formes de proximité (relationnelle, fonctionnelle, etc.) identifiées par Gahinet (2018). Il s'agit, dans ce cadre, d'identifier les types de proximités recherchées par les producteurs et les consommateurs, autrement dit les proximités susceptibles de favoriser l'engagement de ces derniers.

Dans cet article, nous nous proposons ainsi d'analyser les temps de trajet que sont prêts à consentir producteurs agricoles et consommateurs pour se rendre dans des points de vente de produits en CC, définis comme plus ou moins alternatifs. Pour ce faire, nous utilisons les résultats d'une expérience de choix discret menée en 2022 dans le cadre de travaux de thèse (Horvath, 2023). Nous cherchons à identifier les types de proximités recherchées par les producteurs et consommateurs, autrement dit les proximités susceptibles de favoriser l'engagement de ces derniers.

Dans la suite de cet article, une première partie présente une revue de littérature sur l'alternativité des CC. Une deuxième partie présente la méthodologie. Enfin, les résultats des sont présentés et mis en relation, puis discutés dans une dernière partie.

## 2. Revue de la littérature

### 2.1. Les CC, la promesse d'une « autre » économie ?

Quoique les CC ne soient *per se* un mode de commercialisation nouveau[3], ils font l'objet depuis les années 2000, d'un regain d'intérêt face à l'urgence climatique. Les CC sont souvent présentés « *en rupture avec les logiques intensives, spécialisées et intermédiées* » (Velly et Dubuisson-Quellier, 2008). A partir de la littérature, cette partie vise à qualifier les promesses d'alternativité, en nous intéressant au renouvellement des modes de production et de consommation, puis des modes d'échanges commerciaux suscités par les CC.

*2.1.1. Les CC, des modes de production et de consommation alternatifs ?*

Du point de vue de la production agricole d'abord, si les exploitations commercialisant en CC sont très diverses, la littérature montre que nombreuses sont celles présentant des profils alternatifs, peu ou pas intégrés aux filières longues. Elles se caractérisent par de petites structures, adoptant des pratiques productives durables (agriculture raisonnée voire biologique) et portées par de nouveaux profils d'agriculteurs (Guillemin & Margetic, 2022; Morel, 2019). Pour autant, d'autres exploitations, souvent plus grandes, recourent simultanément aux circuits courts et longs (Rouget et al., 2021), s'inscrivant ainsi dans des trajectoires d'hybridation, qualifiées par Sencébé et David (2023) de « *modernisation alternative* ». En effet, les CC présentent pour les producteurs agricoles des intérêts divers articulés autour de processus de diversification multi-dimensionnels : diversification des cultures, des débouchés et des revenus. Dans ce cadre, les CC sont souvent perçus par les producteurs agricoles comme un gage d'autonomie, permettant une meilleure maîtrise des prix et une meilleure valorisation des produits (Sencébé et David, 2023 ; Alonso Ugaglia *et al*., 2020). Du point de vue des consommateurs ensuite, l'achat en CC permet aussi de favoriser, à des degrés divers, les proximités relationnelles entre acteurs de l'alimentation locale. Cela concerne d'abord les relations entre consommateurs et producteurs. Pour les premiers, cette reconnexion peut être synonyme de transfert d'informations, notamment sur la traçabilité du produit et ses caractéristiques propres (Benedetti et al., 2023 ; Roznowicz & Odou, 2021). Dans un contexte de multiplication des scandales sanitaires, une proximité accrue entre producteurs et consommateurs peut rétablir la confiance de ces derniers. Parallèlement pour les seconds, une plus grande proximité avec les consommateurs représente l'opportunité d'échanges directs pouvant permettre une meilleure connaissance des préférences de la demande, mais aussi de sensibiliser aux enjeux et difficultés de l'agriculture et du métier d'agriculteur. Les CC sont également susceptibles de favoriser les proximités entre agriculteurs (qui peuvent se rencontrer sur les marchés de plein vent par exemple), en renouvelant les pratiques d'entraide et en permettant des partages de ressources matérielles (outils, locaux, etc.) et immatérielles (informations plus ou moins stratégiques) (Raimbert et Raton, 2023, 2021; Romeyer, 2012). Ainsi, selon Chiffoleau, la « *capacité transformatrice* » des CC repose notamment sur les échanges entre pairs, permettant de favoriser les innovations durables, en limitant les échecs agronomiques et financiers (Millet, 2015). Aussi, acheter en CC est perçu par les consommateurs comme une manière de soutenir les agriculteurs locaux et de favoriser un meilleur respect de l'environnement (Burchardi et al., 2005 ; Meyerding, 2019 ; Pernin et Petitprêtre, 2013). La promesse de différence d'une « autre » économie (Chiffoleau, 2023) que représentant les CC suppose donc non seulement de nouvelles façons de produire et de consommer, mais aussi de nouvelles façons de commercialiser les denrées alimentaires.

---
[3] Par exemple, la vente ambulante ou à la ferme, les marchés de plein vent.

*2.1.2. Des modes de distribution à alternativité variable*

La vente de produits en CC s'opère à travers une diversité de canaux de distribution plus ou moins nouveaux, mais aussi et surtout, plus ou moins spécialisés et insérés dans des systèmes mondialisés. Ainsi, les GMS vendent elles aussi des produits issus des CC. Parmi les principaux autres canaux de distribution de produits en CC, on compte à la fois des canaux historiques tels que les marchés de plein vent ou la vente à la ferme et d'autres plus innovants, à l'instar des plateformes numériques (Maye, 2013). Cette diversité se traduit également par des niveaux d'alternativité variable.

Ces derniers peuvent se mesurer à l'aune de leur capacité à répondre aux promesses de différences des CC, et tout particulièrement celles relatives, (1) à une rémunération plus juste des agriculteurs et au soutien à l'économie locale et, (2) au développement des proximités relationnelles entre producteurs et consommateurs, et entre producteurs. On trouve dans la littérature plusieurs travaux proposant des typologies des points de vente en CC pouvant éclairer leur niveau d'alternativité.

(1) Roznowicz et Odou (2021) utilisent le cadre théorique des économies de la grandeur (Boltanski et Thévenot, 1991) pour différencier les GMS et les points de vente en CC. Ils définissent les GMS comme appartenant aux modes marchands et industriels, axés sur l'économie d'échelle, tandis que les CC, tels que les AMAP et les marchés, offrent une transparence accrue et favorisent la solidarité entre producteurs et consommateurs, offrant ainsi des produits de qualité à des prix équitables et permettant une prise de décision démocratique.

(2) Quant à Benedetti et al. (2023), ils proposent une classification des points de vente en CC à Nice, basée sur les niveaux de proximité fonctionnelle (comprise dans cette étude comme la présence d'un intermédiaire ou non) et relationnelle. Les AMAP et la vente directe à la ferme offrent les plus hauts niveaux de proximité. Les marchés de plein vent et les épiceries constituent une deuxième classe avec une proximité fonctionnelle élevée mais une perte de connaissances sur l'origine exacte des produits. La dernière classe concerne les commerces utilisant des intermédiaires et la vente en ligne, avec des niveaux de proximité fonctionnelle et relationnelle faibles.

Dans ces études, les points de vente en CC sont donc comparés les uns aux autres selon les différences qu'ils permettent en termes de proximités (relationnelle, fonctionnelle, etc.). Elles mettent en évidence un gradient d'alternativité avec d'un côté, les GMS qui, certes, proposent des produits en CC, mais surtout reproduisent un modèle industriel et standardisé et, de l'autre, des formes de CC particulièrement alternatives, et notamment les formes de vente directe à même de favoriser des liens étroits entre producteurs et consommateurs, voire un engagement logistique de leur part

## 2.2. Mise en relation agriculteurs/consommateurs : des contraintes d'accès qui font émerger des formes d'engagement ?

Les contraintes, notamment logistiques, à la vente et l'achat en CC permettent d'expliquer les difficultés de développement de ces circuits. En effet, les problèmes relatifs à la rencontre entre offre et demande pour les produits vendus en CC recèlent à la fois des questions d'adéquation, mais aussi et surtout et de rencontre (physique) entre l'offre et la demande. En effet, tandis que les consommateurs tendent à préférer, pour leurs achats alimentaires, des points de vente proposant une grande largeur de gamme (Crédoc, 2017) – incluant par exemple, produits exotiques, hors saison (bananes, tomates, etc.) ou produits non

alimentaires – et localisés à proximité du domicile/lieu de travail, les points de vente en CC peuvent offrir des gammes plus restreintes, avec des approvisionnements plus ou moins irréguliers (volumes plus faibles, saisonnalité des produits). Ils sont, par ailleurs, moins nombreux et plus dispersés.

Ainsi, la fragmentation des flux des chaînes courtes (Raton et al. 2020) suppose, le plus souvent, une multiplication des débouchés pour les producteurs agricoles et, parallèlement, des lieux d'approvisionnement alimentaire pour les consommateurs, aussi constatée par (Essers and Poulot, 2019) qui notent qu'acheter en CC est souvent synonyme de fragmentation des lieux alimentaires fréquentés, entre supermarchés et points de vente en CC. Dès lors, recourir aux CC exige des changements de pratiques et d'organisation qui ne relèvent pas seulement des habitudes de production et de consommation, mais aussi des habitudes de mobilité (Raton et Raimbert, 2019). En effet, notons que le terme de circuits courts met, d'abord, en avant une différence de nature logistique : le raccourcissement des chaînes d'approvisionnement, induisant une réduction du nombre d'acteurs le long de cette chaîne. Certains auteurs mettent ainsi en exergue l'importance de la problématique spatiale de la rencontre entre offre et demande des produits en CC dans l'analyse des dynamiques de transition des systèmes agri-alimentaires vers plus de durabilité (Horvath, 2023). On peut dès lors considérer que le transport nécessaire pour vendre ou acheter des produits en CC peut entrer dans les arbitrages réalisés par les producteurs et les consommateurs.

Ces contraintes, et leur surpassement, peuvent induire une forme d'engagement des producteurs et des consommateurs pour vendre et acheter dans des points de vente en CC alternatifs. Selon Amilien et al. (2022), dans une analyse portant sur les fondements de la consommation de produits de qualité, l'engagement comporte deux dimensions : la première est la dimension morale ; l'engagement se fonde alors sur un processus de délibération permettant de déterminer si les promesses de différence des produits sont à la hauteur des attentes des consommateurs ; la seconde correspond à l'engagement physique, soit le fait d'accepter de se rendre dans tel ou tel lieu, en l'occurrence dans notre cas dans tel ou tel type de point de vente. Ainsi, dans un contexte où le manque de temps fait émerger de nombreux services commerciaux ou logistiques (service de livraison des achats alimentaires, service de préparation et livraison de plats etc.), le temps de trajet représente une forme d'engagement des agriculteurs et des consommateurs à s'impliquer dans la logistique des CC ; ces trajets effectués représentant une distance à parcourir (temps de trajet) à laquelle est associée un coût (coût de transport).

D'autres travaux mettent en évidence l'influence des proximités sur l'engagement des acteurs en CC. Ainsi, dans une étude portant sur les motivations des consommateurs en faveur des AMAP, Dufeu et Ferrandi (2011) analysent les différentes formes de proximité permises par les AMAP et comment ces dernières influent sur la confiance des consommateurs, et partant, sur leur satisfaction et leur engagement. Ils distinguent 4 types de proximités : la proximité identitaire indique que les valeurs des consommateurs sont proches de celles de l'AMAP ; la proximité relationnelle correspond à la présence de liens entre agriculteurs et consommateurs ; la proximité de processus indique que le consommateur connaît les pratiques de production des agriculteurs, et la proximité d'accès que l'AMAP est proche géographiquement des consommateurs. Nous faisons ici l'hypothèse que cette approche par les proximités pourrait s'appliquer à d'autres points de vente d'une part et aux agriculteurs et leur choix de débouchés, d'autre part aux consommateurs et à leurs choix d'un point de vente où acheter en CC.

Nous pouvons considérer que la proximité qui est recherchée par les consommateurs et les agriculteurs, pourrait se traduire par un engagement physique accru, mesurable par leur propension à parcourir plus de kilomètres/faire des trajets plus longs pour accéder à un point de vente plus ou moins alternatif, en tenant compte du prix d'achat ou de vente. Cette approche permet ainsi une analyse plus fine des préférences des consommateurs et des producteurs, ainsi que de la rencontre entre offre et demande, en prenant en compte le coût du temps de trajet pour les CC. Cette approche semble adaptée pour prendre en compte le fait que le marché est en cours de structuration et que des disparités territoriales existent. Pour mesurer l'engagement logistique des consommateurs et agriculteurs en CC, nous nous demandons à la fois (1) si le niveau d'alternativité des points de vente influence les temps de trajet que sont prêts à consentir les agriculteurs et les consommateurs pour vendre ou acheter en CC et (2) quelles sont les autres formes de proximités susceptibles de contrebalancer l'importance de la proximité spatiale et par là même de favoriser l'engagement logistique des agriculteurs comme des consommateurs.

## 3. Méthodologie et données
### 3.1. Données mobilisées

Les données utilisées dans cet article CC proviennent de deux expériences de choix discrets menées en 2022 auprès de 1022 consommateurs et 154 maraîchers en France métropolitaine, dans le cadre du travail de thèse de Horvath (2023). La base de données contient les expériences de choix discrets, des informations sur les comportements d'achats en circuits courts et les perceptions de ces circuits par les consommateurs, et des informations sur les pratiques de vente et les perceptions de ces circuits par les maraîchers. Les questions sur les perceptions sont présentées sous la forme d'échelles de Likert.

Afin de calculer les consentements à payer (CAP) et à vendre (CAV) dans différents points de vente plus ou moins alternatifs, nous mobilisons les questionnaires de cette thèse, administrés en ligne. Les sondés, maraîchers comme consommateurs, ont choisi parmi 6 propositions, dont une option de ne pas choisir. L'expérience a été répétée 6 fois par sondés. Ces choix nous permettent d'estimer les paramètres d'utilité, qui représentent les préférences pour chaque attribut (Batsell et Louviere, 1991) et ainsi de calculer les CAP et CAV. Les attributs et leurs niveaux ont été distribués aléatoirement dans les scénarios. Les préférences analysées sont : (1) pour les consommateurs, l'achat d'un panier de 4 kg de légumes, sans spécification du type de légumes présents ; (2) pour les agriculteurs, la vente d'un surplus de production de 100 kg. Les attributs utilisés sont : (1) pour les consommateurs, le type de point de vente, le prix du panier, le temps de trajet pour aller dans le point de vente, si les produits du panier sont bio ou non, et si des évènements à la ferme sont organisés par les maraîchers ou non ; (2) pour les maraîchers, le type de point de vente, le prix par kg, le temps de trajet pour livrer le point de vente, s'il existe de l'entraide entre agriculteurs pour livrer le point de vente ou non, et si les consommateurs sont intéressés par l'organisation d'évènements à la ferme ou non. Les cartes de choix sont disponibles en Annexe 1.

Afin d'étudier les effets des perceptions sur les CAP et CAV, nous mobilisons les données de perceptions des CC, proposées sous la forme d'échelles de Likert allant de 1 à 5. Pour les consommateurs, nous avons utilisé les perceptions suivantes : « Par conviction, je n'achèterais pas un produit des circuits courts dans une grande surface » et « Les circuits courts permettent de soutenir les agriculteurs ». Pour les maraîchers, nous avons utilisé l'échelle de réponse à la question « Je préfère vendre en circuits courts qu'en circuits longs », qui peut indiquer un engagement physique des agriculteurs à la vente en CC.

### 3.2. Mesurer l'alternativité des points de vente à partir des proximités

Pour mesurer l'alternativité de ces points de vente, nous les qualifions en nous basant sur les proximités qu'ils permettent. Pour cela, nous reprenons les types de proximité utilisées par Dufeu et Ferrandi (2011), en considérant l'alternativité des points de vente à l'aune de leur capacité à favoriser une reconnexion entre agriculteurs et consommateurs, que nous traduisons par la production de :

(1) proximités identitaires, soit, dans notre étude une meilleure rémunération aux agriculteurs. Elle est mesurée en fonction de deux critères : (a) le prix de la vente est directement ou non versé à l'agriculteur, supposant une répartition plus ou moins équitable de la valeur le long de la chaîne ; (b) la vente et l'achat des produits fait l'objet ou non d'une contractualisation, supposant une plus ou moins équitable répartition du risque.

(2) proximités relationnelles, se traduisant, dans notre étude, par (a) la possibilité ou non d'échanger directement entre producteurs et consommateurs, mais également (b) la qualité de ces échanges (mesurée par la disponibilité variable des agriculteurs en fonction des points de vente).

Cette classification, qui repose sur la description des points de vente proposées dans les enquêtes, est montrée sur le Tableau 1.

*Tableau 1: Description des points de vente proposés dans les enquêtes en fonction d'indicateurs d'alternativité.*

| Type de point de vente | Type de proximité | Description correspondante des points de vente dans l'enquête | Niveau de proximités |
|---|---|---|---|
| Association de consommateurs | Identitaire | Prix payé versé directement à l'agriculteur ; Adhésion des consommateurs pendant 3 mois | Fort |
| | Relationnelle/ Processus | Possibilité d'échanges entre agriculteurs et consommateurs ; disponibilité importante des agriculteurs | Fort |
| Ferme | Identitaire | Prix payé versé directement à l'agriculteur ; pas de contrat | Moyen |
| | Relationnelle/ Processus | Possibilité d'échanges entre agriculteurs et consommateurs ; disponibilité importante des agriculteurs | Fort |
| Marché | Identitaire | Prix payé versé directement à l'agriculteur ; pas de contrat | Moyen |
| | Relationnelle/ Processus | Possibilité d'échanges entre agriculteurs et consommateurs, selon l'affluence sur le marché. La qualité des échanges n'est pas garantie | Moyen |
| Drive | Identitaire | Prix payé versé directement à l'agriculteur | Moyen |
| | Relationnelle/ Processus | Possibilité d'échanges entre agriculteurs et consommateurs uniquement lorsqu'ils se croisent. La présence et la qualité des échanges ne sont pas garanties. | Faible |
| Supermarché | Identitaire | Le prix payé va à l'intermédiaire, qui en réserve une partie à l'agriculteur | Faible |
| | Relationnelle/ Processus | Pas de possibilité d'échanges entre agriculteurs et consommateurs | Faible |

*Légende : Pour la proximité identitaire : Forte = 2 critères présents, Moyenne = 1 critère présente, Faible : aucun critère ; pour la proximité relationnelle : Forte = présence et qualité garantie, Moyenne = présence garantie, qualité non garantie, Faible : présence et qualité non garantie, voire impossible.*

En plus du choix d'un point de vente, nous utilisons aussi d'autres variables proposées aux sondés pour évaluer le rôle des différentes formes de proximité. Ces variables sont aléatoirement attribuées aux points de vente et nous permettent de mesurer de façon plus précise le rôle de différentes formes de proximité : relationnelles (événement à la ferme et entraide) ;

de processus (bio) et fonctionnelle (largeur de gamme). La variable sur la volonté d'organiser des évènements à la ferme, exprimant une volonté de rencontre entre agriculteurs et consommateurs (voir Annexe 1), permet d'évaluer la valorisation de la proximité relationnelle faite par les consommateurs et les agriculteurs. La variable sur l'entraide entre agriculteurs indique aussi une valorisation de la proximité relationnelle faite, cette fois, entre les agriculteurs. Le fait pour les consommateurs d'accepter de payer plus cher pour des produits bio peut indiquer une certaine valorisation de la proximité de processus, définie comme l'importance qu'accorde le consommateur au « fonctionnement interne du magasin (ses fournisseurs, ses produits, sa logistique...) qui sera garant de la qualité des produits ou du service attendus. » (Bergadaà et Del Bucchia (2009). Gahinet (2018) note que en adaptant cette définition aux CC, elle devient le « partage de connaissances sur le fonctionnement interne du CC, apprécié à travers les connaissances sur l'origine des produits et les méthodes de production et de distribution » (Dufeu and Ferrandi, 2011.; Hérault-Fournier, 2013; Hérault-Fournier et al., 2012). Le choix d'une largeur de gamme importante pour les consommateurs peut aussi indiquer une valorisation d'une forme de proximité fonctionnelle, définie comme la « commodité qu'offre le magasin en termes d'efficacité pour réaliser ses courses alimentaires » (Gahinet, 2018).

### 3.3. Profil des enquêtés

Les statistiques descriptives des deux échantillons est présenté dans le Tableau 2. Elles montrent que l'échantillon est proche de la population réelle pour les consommateurs. Les femmes sont plus représentées que la moyenne, mais cela peut être expliqué par le fait qu'elles restent à près de 60% les principales responsables d'achat dans les foyers (sondage Ipsos, 2019). Les employés sont aussi plus représentés que dans la population française (voir Figure 11) ce qui peut être en partie expliqué par le fait que les femmes sont plus représentées dans l'échantillon, et que la CSP « Employés » est la principale pour les femmes en 2021 (pour 41,1 % d'entre elles, d'après l'INSEE en 2021, contre 12 % des hommes).

Afin d'avoir une plus grande homogénéité dans les statistiques descriptives et de limiter le biais dans les choix effectués, nous ne retenons que les résultats pour les 154 agriculteurs sondés ayant répondu avoir une activité de maraîchage dans leur exploitation agricole. Les maraîchers ayant répondu sont plus enclins à vendre en circuits courts : 35 % vendent uniquement en CC, et 90 % d'entre eux indiquent utiliser au moins un débouché en CC. Les agriculteurs vendant en CC, et qui plus est comme seul circuit, sont surreprésentés dans cette enquête par rapport au contexte national actuel (23% des exploitations françaises vendent au moins un type de produit en CC, Agreste, RA, 2020). De même, les exploitations produisant sous label bio représentent 60 % de l'échantillon, alors que la moyenne nationale est de 12 % en 2020 toute OTEX confondue (RA, 2020), et était de 13 % pour les exploitations maraîchères en 2010 (Agreste, 2010). Deux tiers des exploitations ne sont pas orientées uniquement en maraîchage, et 14% d'entre elles ne vendent qu'en circuits longs.

*Tableau 2 : Statistiques descriptives des maraîchers et des consommateurs sondés*

| | Maraîchers | Consommateurs |
|---|---|---|

|  | Moyenne | Ecart-type | Moyenne | Ecart-type | **Moyenne nationale (2022)** |
|---|---|---|---|---|---|
| **Variables sociodémographiques** | | | | | |
| Femmes (%) | 36 | 23 | 66 | 24 | 52 |
| Age (années) | 41 | 11 | 43 | 15 | 42 |
| Diplôme supérieur hors agriculture (%) | 43 | 25 | 47 | 25 | 40 |
| Actifs (%) | | | 67,45 | 22 | 71 |
| Revenus du foyer (€/personne/mois) | | | 2642 | 1096 | 1837 |
| Retraités (%) | | | 14 | 12 | 17 |
| **Variables agricoles** | | | | | |
| Durée installation (années) | 11 | 11 | | | nc |
| SAU maraîchage (ha) | 6,58 | 10,94 | | | 10,52 |
| SAU totale (ha) | 41,56 | 67,3 | | | 64,55 (toute OTEX confondue) |
| Part maraîchage SAU (%) | 43,63 | 38,27 | | | nc |
| Bio (%) | 60 | 24 | | | 12 (toute OTEX confondue) |
| 100% CC (%) | 35 | 23 | | | nc |
| 100% C Longs (%) | 14 | 12 | | | nc |
| 100% maraîchage | 31 | 22 | | | nc |
| EBE/ha (€/ha) | 4786 | 4997 | | | 2500 (conventionnel en 2013) / 3300 (agriculture biologique en 2013) |
| Effectifs | | 154 | | 1022 | |

*Source : Horvath, 2023*

*Légende : SAU (Surface Agricole Utile), CC (Circuits courts), C Longs (Circuits longs), EBE (Excédent Brut d'Exploitation)*

*Nc : en l'absence de statistiques du Recensement Agricole 2020*

En s'intéressant aux perceptions des consommateurs et des agriculteurs, présentées dans le Tableau 3, nous pouvons voir que plus de la moitié des agriculteurs sont d'accord avec le fait qu'ils préfèrent vendre en CC plutôt qu'en circuits longs. Cela rejoint le fait que la plupart d'entre eux a déjà développé au moins une activité en circuit court.

*Tableau 3: Accord des sondés avec les perceptions proposées dans l'étude*

| **Sondés** | **Perception** | **Non, pas du tout** | **Plutôt non** | **Ni oui ni non** | **Plutôt oui** | **Oui, totalement** |
|---|---|---|---|---|---|---|
| **Maraîchers** | « Je préfère vendre en CC qu'en circuits longs » | 6,20% | 5,47% | 17,88% | 21,17% | 49,27% |
| **Consommateurs** | « Je n'achèterais pas des produits des CC dans les supermarchés » | 24,45% | 26,64% | 33,75% | 9,12% | 8,03% |
| | « Les CC permettent de soutenir les agriculteurs locaux » | 2,19% | 4,38% | 16,42% | 39,78% | 37,23% |

*Source : Horvath, 2023*

Pour les consommateurs, la phrase « *Les CC permettent de soutenir les agriculteurs locaux* » est une perception à laquelle adhèrent les ¾ des consommateurs sondés. La perception « Je n'achèterais pas des produits des CC dans les supermarchés », a l'effet inverse : la moitié d'entre eux ne sont pas d'accord avec l'affirmation, (1/3 n'a pas d'avis, et 15% d'accord).

### 3.4 Construction de sous-groupes

Nous définissons des sous-groupes pour identifier les consommateurs et les agriculteurs que nous considérons les plus engagés dans les CC. Pour les consommateurs, nous proposons 4 sous-groupes permettant de distinguer ceux qui achètent déjà en CC (groupe A), et ceux qui n'achètent jamais en CC (groupe B). Nous distinguons aussi les consommateurs en accord avec l'affirmation- « *Je préfère ne pas acheter en CC au supermarché* » (groupe C) de ceux qui sont en accord avec « *Les CC permettent de soutenir les agriculteurs* » (groupe D), ces groupes ne se chevauchant pas. Pour les maraîchers, nous distinguons 3 sous-groupes : ceux qui commercialisent uniquement en CC (groupe 1), ceux qui ne commercialisent qu'en CL (groupe 2), ceux qui commercialisent à la fois en CC et en CL (groupe 3) et ceux qui sont en accord avec la préférence « *Je préfère ventre en CC qu'en CL* » (groupe 4). Les statistiques descriptives par sous-groupes sont présentées en Annexe 2.

### 3.5. Modélisation

Durant les EPD, les sondés devaient choisir entre différentes alternatives avec différents niveaux d'attributs. En faisant l'hypothèse que les préférences sont distribuées aléatoirement entre les individus, un modèle économétrique d'utilité aléatoire est justifié. De plus, les sondés sont aussi caractérisés par une hétérogénéité inobservable, qui peut être contrôlée en utilisant des coefficients aléatoires. Les utilités suivantes ont été estimées avec des modèles. La notation $U_j(X_i)$ correspond à l'utilité de l'individu *i* pour le point de vente *j*.

$$U_j(X_i) = e_{ij} + \beta_i X_j \qquad (1)$$

Où pour les consommateurs :

$$U_j(X_i) = \alpha_{ij}^C + \beta_{P_i}^C * P_j + \beta_{G_i}^C * G_j + \beta_{T_i}^C * T_j + \beta_{B_i}^C * B_j + \beta_{L_i}^C * L_j + e_{ij} \qquad (2)$$

La variable $P_j$ est continue correspondant au prix du panier, $G_j$ est une variable dummy indiquant la largeur de gamme dans le point de vente, $T_j$ est une variable continue indiquant le temps de trajet pour aller dans le point de vente, $B_j$ est une variable dummy indiquant si les légumes sont vendus sous label d'agriculture biologique ou non, $L_j$ est une variable dummy indiquant si les agriculteurs proposent d'organiser des évènements sur leur ferme ou non. Le coefficient $\alpha_{ij}^C$ correspond à la préférence intrinsèque du consommateur $i$ pour le débouché $j$.

Pour les maraîchers, l'utilité est sous la forme suivante :

$$U_j(X_i) = \alpha_{ij}^P + \beta_{P_i}^P * P_j + \beta_{T_i}^P * T_j + \beta_{E_i}^P * E_j + \beta_{L_i}^P * L_j + e_{ij} \qquad (3)$$

La variable $P_j$ est continue et correspond au prix pour 1 kg de légumes vendus dans le point de vente j. La variable $E_j$ est une variable dummy indiquant l'entraide, si d'autres agriculteurs vendent déjà dans le point de vente et sont prêts à aider. La variables $T_j$ est une variable continue, indiquant le temps de trajet en minute pour aller livrer le point de vente. La variable $L_j$ est une variable dummy qui prend 1 quand les consommateurs veulent organiser des évènements sur la ferme. Le coefficient $\alpha_{ij}^P$ correspond à la préférence intrinsèque du maraîcher $i$ pour le débouché $j$.

Avec ces utilités agrégées, nous calculons les CAP et CAV marginaux pour chaque attribut $k$ :

$$CAP_k = -\frac{\beta_k^C}{\beta_P^C} \qquad (4)$$

$$CAV_k = -\frac{\beta_k^P}{\beta_P^P} \qquad (5)$$

Les données sont analysées en utilisant le package Apollo du logiciel R (Hess et Palma, 2022). Des modèles logit multinomiaux sont d'abord calculés. Le prix devrait avoir un effet positif pour les maraîchers, négatif pour les consommateurs. Nous nous attendons à ce que les variables binaires aient un effet positif quand elles sont présentes (entraide, évènements, bio), et que les consommateurs recherchent une largeur de gamme importante.

Nous calculons ensuite des mixed logit pour prendre en compte l'hétérogénéité entre les individus. Les logit mixed sont utilisés avec 500 sobol draws. Les logit multinomiaux et les mixed logit prennent en considération la dimension de panel des données. Nous considérons que toutes les variables sauf le prix sont des paramètres aléatoires, distribués selon des gaussiennes.

### 3.6. Délimitation de potentiels de marché et lien avec l'engagement

Nous estimons des potentiels de marché pour les différents points de vente en CC, en étudiant le CAP des consommateurs et le CAV des agriculteurs en fonction des temps de trajet qu'ils et donc de la valorisation du temps de trajet que les expériences de choix révèlent. Cela permet de mettre en avant les arbitrages réalisés par les consommateurs et les agriculteurs entre prix et temps de trajets, et surtout de réaliser des comparaisons entre les points de vente, en particulier entre les plus alternatifs et les moins alternatifs.

Afin d'étudier les potentiels de marché pour les différents points de vente en CC proposés dans l'enquête, des CAP « généralisées » sont calculées. En effet, certains des canaux diffèrent structurellement des autres au regard de leurs caractéristiques. Par exemple, le temps de trajet des maraîchers en vente directe à la ferme est (toujours) nulle ; au supermarché, les consommateurs trouveront toujours d'autres produits que des légumes, la largeur de gamme est toujours la plus grande. Partant de ce constat, nous posons :

$$CAPG_j = \frac{\alpha_j^C + \beta_T^C \hat{T}_j + \beta_j^C \hat{X}_J}{\beta_P^C} \quad (6)$$

De la même manière, nous pouvons obtenir un CAV généralisé :

$$CAVG_j = \frac{\alpha_j^P + \beta_T^P \hat{T}_j + \beta_j^P \hat{X}_J}{\beta_P^P} \quad (7)$$

Où $\hat{T}_j$ est la durée moyenne que mettent les consommateurs (maraîchers) pour aller acheter (vendre) les légumes via le canal $j$. Et idem pour $\hat{X}_J$, correspond à la moyenne des autres variables.

Afin d'isoler et de mettre en avant l'influence des temps de trajet sur les préférences des acteurs en faveur des différents débouchés, nous pouvons réécrire l'équation 8 de la manière suivante :

$$CAPG_j(d_j) = \frac{\alpha_j^C + \beta_j^C \hat{X}_j}{\beta_P^C} + \frac{\beta_{Tj}^C \widehat{T}_j}{\beta_P^C}$$

$$= n_j^C + b_{T_j}^C \hat{T}_J \quad (8)$$

De la même manière, nous obtenons le CAVG suivant :

$$CAVG_j(d_j) = \frac{\alpha_j^P + \beta_T^P \hat{X}_j}{\beta_P^P} + \frac{\beta_{Tj}^P \widehat{X_T}}{\beta_P^P}$$

$$= n_j^P + b_{T_j}^P \widehat{T_j} \quad (9)$$

On note : $\Delta CAPG_j = CAPG_j - CAVG_j$. Nous faisons l'hypothèse qu'un point de vente est d'autant plus privilégié que $\Delta CAPG$, et donc l'écart entre CAPG et CAVG, est grand.

A l'aide des deux précédentes équations, nous traçons des droites permettant d'approcher les temps de trajets et prix pour lesquels les consommateurs seraient prêts à payer, les maraîchers à vendre. Cela nous permet de prendre en compte le ou les types de proximités permises par le point de vente, et de voir les différences d'engagement, à travers les temps de trajet et la proximité spatiale.

## 4. Résultats

### 4.1. Consentement à vendre, consentement à payer

*4.1.1. Consommateurs*

Les résultats des modèles MNL sont présentés en Annexe 3. Les CAP qui en découlent sont présentés dans le Tableau 4. Si nous regardons les préférences intrinsèques pour les points de vente, nous pouvons voir que tous les consommateurs préfèrent la ferme, à l'exception des non-acheteurs en CC (groupe B), qui ont le CAP le plus élevé pour l'achat en CC au supermarché. Ensuite, les consommateurs ont des CAP élevés pour le drive, puis pour supermarché et le marché. L'association de consommateurs arrive pour tous en dernière position, avec les CAP les plus faibles. Cela peut s'expliquer par la contrainte de contractualisation financière dans le temps, qui a pu représenter un frein pour les sondés.

Si l'on s'intéresse aux temps de trajet, nous pouvons voir que le CAP diminue significativement avec chaque minute de temps de trajet. Cela montre l'importance de prendre en compte le transport lorsque l'on s'intéresse à l'achat local. Ce constat est tempéré pour l'association de consommateurs où le CAP diminue le plus faiblement avec chaque minute de trajet . Cela suggère que l'association a été le moins choisie par les sondés, mais que ceux qui l'ont choisie l'on fait sans porter une grande importance aux temps de trajet. A l'inverse, bien que la ferme soit le point de vente préféré des consommateurs en général, c'est pour ce point de vente, ainsi que pour le drive, que le CAP des consommateurs diminue le plus fortement avec chaque minute de trajet.

Les sondés ont tous valorisé les différentes formes de proximités proposées dans l'étude, les CAP étant toujours significatifs et positifs pour l'organisation d'évènements à la ferme (proximité relationnelle), pour l'achat en agriculture biologique (proximité de processus) et pour la largeur de gamme (proximité fonctionnelle).

Les consommateurs déjà acheteurs des CC (groupe A) ont des CAP plus élevés que la moyenne. Ce CAP est néanmoins réduit plus fortement que la moyenne avec chaque minute supplémentaire de temps de trajet. Ils valorisent aussi plus les différentes formes de proximités proposées dans l'étude. Au contraire, les non-acheteurs des CC (groupe B) ont les CAP les plus faibles. Ils restent néanmoins intéressés par la largeur de gamme, l'achat bio et la possibilité de créer du lien avec les agriculteurs. Les consommateurs déclarant préférer ne pas acheter en CC au supermarché (groupe C) ont les CAP les plus élevés, mais sanctionnent aussi plus le temps de trajet, hormis pour l'association de consommateurs. Les consommateurs liant achat en CC avec soutien des agriculteurs locaux (groupe D) ont des CAP légèrement plus élevés que la moyenne pour tous les points de vente, et similaires à la moyenne concernant les temps de trajet et les formes de proximité proposées dans l'étude.

|  | Tous | CC (A) | CL (B) | Préf.1 (C) | Préf.2 (D) |
|---|---|---|---|---|---|
| Ferme | 6,013*** (0,94) | 7,512*** (1,751) | 5,003*** (1,029) | 11,378*** (8,71) | 6,29*** (1,054) |
| Marché | 5,56*** (0,92) | 6,998*** (1,592) | 4,653*** (1,124) | 9,649*** (7,021) | 5,843*** (1,005) |
| Supermarché | 5,63*** (0,98) | 6,78*** (1,646) | 5,128*** (1,251) | 9,219*** (7,147) | 5,874*** (1,073) |
| Drive | 5,69*** (0,85) | 7,194*** (1,52) | 4,706*** (1,035) | 10,566*** (7,256) | 6,051*** (0,943) |
| Association | 4,21*** (0,98) | 5,632*** (1,59) | 3,277*** (1,327) | 7,096*** (6,112) | 4,58*** (1,11) |
| TT Ferme | -0,103*** (0,041) | -0,104*** (0,059) | -0,104*** (0,057) | -0,208*** (0,234) | -0,097*** (0,042) |
| TT Marché | -0,076*** (0,033) | -0,081*** (0,049) | -0,067*** (0,041) | -0,119*** (0,183) | -0,077*** (0,035) |
| TT Supermarché | -0,070*** (0,070) | -0,080*** (0,046) | -0,064*** (0,037) | -0,117*** (0,178) | -0,069*** (0,031) |
| TT Drive | -0,133*** (0,07) | -0,152*** (0,064) | -0,11*** (0,061) | -0,25*** (0,236) | -0,134*** (0,046) |
| TT Asso | -0,050*** (0,058) | -0,063*** (0,083) | -0,031* (0,079) | -0,004 (0,276) | -0,061*** (0,065) |
| Organisation d'évènements | 0,257*** (0,162) | 0,334*** (0,247) | 0,165*** (0,209) | 0,515*** (0,932) | 0,25*** (0,174) |
| Bio | 0,600*** (0,231) | 0,723*** (0,346) | 0,44*** (0,298) | 1,294*** (1,33) | 0,588*** (0,24) |

|  | Tous | CC (A) | CL (B) | Préf.1 (C) | Préf.2 (D) |
|---|---|---|---|---|---|
| Gamme | 0,341*** | 0,407*** | 0,232*** | 0,643*** | 0,381*** |
|  | (0,238) | (0,335) | (0,342) | (1,14) | (0,252) |
| Nombre obs. | 6126 | 3630 | 2496 | 1116 | 4662 |

*Tableau 4 : Consentement à payer des consommateurs pour 1 kg de légumes vendus en circuits courts. CC : acheteurs des CC, CL : non-acheteurs des CC, Préf.1 : préférence « je préfère ne pas acheter des produits en CC au supermarché », Préf.2 : préférence : « Les CC permettent de soutenir les agriculteurs locaux ». Source : autrices.*

*4.2.2. Maraîchers*

Les résultats des modèles MNL pour les maraîchers sont présentés en Annexe 4. Les CAV sont présentés dans le Tableau 5. Le CAV se lit inversement au CAP. Plus le CAP augmente, plus les consommateurs apprécient l'attribut. A l'inverse, plus le CAV augmente, moins les maraîchers apprécient l'attribut. Si le CAV pour un point de vente est faible, ou non-significatif, cela veut dire que les maraîchers acceptent de donner le surplus dans le point de vente. Si le CAV pour un point de vente est élevé, cela indique au contraire que les maraîchers n'acceptent de vendre que si les prix sont élevés.

Nous pouvons voir sur l'Annexe 4 que les sondés apprécient globalement moins le drive, probablement à cause des contraintes logistiques qui lui étaient liées en description *(« Les paniers doivent être livrés dans un délai maximal de 2 jours ouvrés après la commande. Il est possible de livrer du mardi au samedi de 16h à 19h. Il est possible de coupler la livraison à un autre trajet »*, voir Annexe 1). Ceux qui ne vendent qu'en circuits longs (groupe 2) sont les seuls à avoir un CAV positif et significatif pour la vente à la ferme. Les temps de trajet ne sont globalement pas significatifs, à part pour l'association de consommateurs. Les agriculteurs ayant tendance à sous-estimer leurs coûts logistiques, la distinction des points de vente par temps de trajet semble peu significative pour les agriculteurs.

Nous proposons alors d'étudier les résultats des modèles sans cette distinction pour les maraîchers. Le modèle MNL est proposé ci-dessous.

|  | Tous | CC (1) | CL (2) | CC/CL (3) | Préf.CC (4) |
|---|---|---|---|---|---|
| Ferme | -1,92 | -3,065 | 6,84*** | -2,89 | -2,82 |
|  | (1,94) | (4,74) | (3,11) | (2,80) | (2,68) |
| Marché | 2,43** | 5,74** | 4,057* | 0,56 | 3,15* |
|  | (1,47) | (3,45) | (2,85) | (2,16) | (1,85) |
| Supermarché | 1,87** | 4,12* | 2,78 | 0,47 | 3,48* |

|  | Tous | CC (1) | CL (2) | CC/CL (3) | Préf.CC (4) |
|---|---|---|---|---|---|
|  | (1,59) | (3,02) | (2,96) | (2,51) | (1,88) |
| Drive | 3,18*** | 6,76** | 3,08 | 1,55 | 4,16*** |
|  | (1,45) | (3,56) | (2,89) | (2,19) | (1,81) |
| Association | 0,34* | -0,58 | 4,26* | 0,035 | 0,39 |
|  | (1,63) | (3,78) | (2,71) | (2,31) | (2,07) |
| TT | 0,049*** | 0,030 | 0,085. | 0,043. | 0,041. |
|  | (0,024) | (0,055) | (0,068) | (0,030) | (0,031) |
| Organisation d'évènements | 0,36 | -0,72 | 0,85 | 0,87. | 0,46 |
|  | (0,36) | (0,80) | (1,01) | (0,53) | (0,46) |
| Entraide | -0,99*** | -1,002 | -00052 | -1,30*** | -1,62** |
|  | (0,44) | (1,03) | (0,50) | (0,65) | (0,75) |
| Nombre obs. | 924 | 324 | 90 | 468 | 630 |

*Tableau 5 : CAV des maraîchers (CC : uniquement en CC, CL : uniquement en CL, CC/CL : utilisant les deux types de débouchés, Préf.CC : préférant vendre en CC qu'en circuits longs)*

Nous pouvons voir que les maraîchers ne commercialisant qu'en CC (groupe 1) ont des CAV plus élevés, mais que celui-ci est indifférent aux temps de trajet. Les agriculteurs ne commercialisant qu'en circuits longs (groupe 2) acceptent de donner le surplus au supermarché et au drive, et apprécient moins le marché et l'association. Ce sont également ceux qui sanctionnent le temps de trajet le plus fortement. Ceux qui utilisent les deux débouchés (groupe 3) semblent indifférents au point de vente, mais sont très intéressés par l'entraide. Ceux qui ont déclaré préférer vendre en CC (groupe 4) ont un CAV non-significatif pour l'association, ce qui indique qu'ils sont prêts à donner le surplus dans ce point de vente. Ils ont sinon des CAV plus élevés pour le marché, le supermarché et le drive. Ce sont eux qui valorisent le plus l'entraide (diminution du CAV de 1€62 en présence de celle-ci).

### 5.2. Potentiels de marché

#### 5.2.1. Tous les sondés (maraîchers et consommateurs)

Les faibles significativités des temps de trajet par point de vente des sous-groupes maraîchers, et de leurs préférences pour les points de vente, indiquent une certaine homogénéité des maraîchers ayant répondu à l'enquête. De ce fait, dans cette section, nous n'utilisons que les données de tous les maraîchers regroupés.

Si l'on s'intéresse aux écarts entre le CAP généralisé (CAPG) et le CAV généralisé (CAVG), nous obtenons les courbes de la Figure 1. Ces courbes permettent de représenter les arbitrages entre les préférences intrinsèques pour le point de vente, et le temps de trajet pour y acheter les produits, ou les livrer. Par exemple, sur la courbe du marché de la Figure 1, les consommateurs ont un CAP global supérieur au CAV global des agriculteurs, ce qui indique un potentiel de marché, jusqu'à un temps de trajet de 30 minutes. Au-dessus de 30 minutes, le CAPG est inférieur au CAVG.

Les courbes de la Figure 1 montrent un fort potentiel de marché pour la ferme, grâce aux agriculteurs acceptant de donner leur surplus, et l'absence de temps de livraison pour les agriculteurs. Le supermarché et le marché présentent aussi un potentiel de marché important, les consommateurs et maraîchers sondés acceptant des temps de trajet d'à peu près 35min, pour un prix autours de 4€/kg. Pour l'association de consommateurs, les temps de trajets permettant un écart positif entre CAPG et CAVG vont jusqu'à plus de 40min, pour un prix de 2,5€/kg. Ce potentiel de marché est dû au fait que les agriculteurs apprécient ce point de vente, et acceptent d'y vendre à faible prix. Le drive a le potentiel le plus faible (15 min, pour 4€/kg), les consommateurs refusant de longs trajets, et les maraîchers ayant des CAVG élevés pour ce point de vente.

Nous pouvons ainsi voir que pour les points de vente les plus alternatifs (association de consommateurs, ferme), le potentiel de marché est plus grand, car les maraîchers acceptent de vendre à de faibles prix, et pour l'association, les consommateurs acceptent des temps de trajets plus élevés. Le drive, qui faisait partie des points de vente les moins alternatifs, a le potentiel de marché le plus faible.

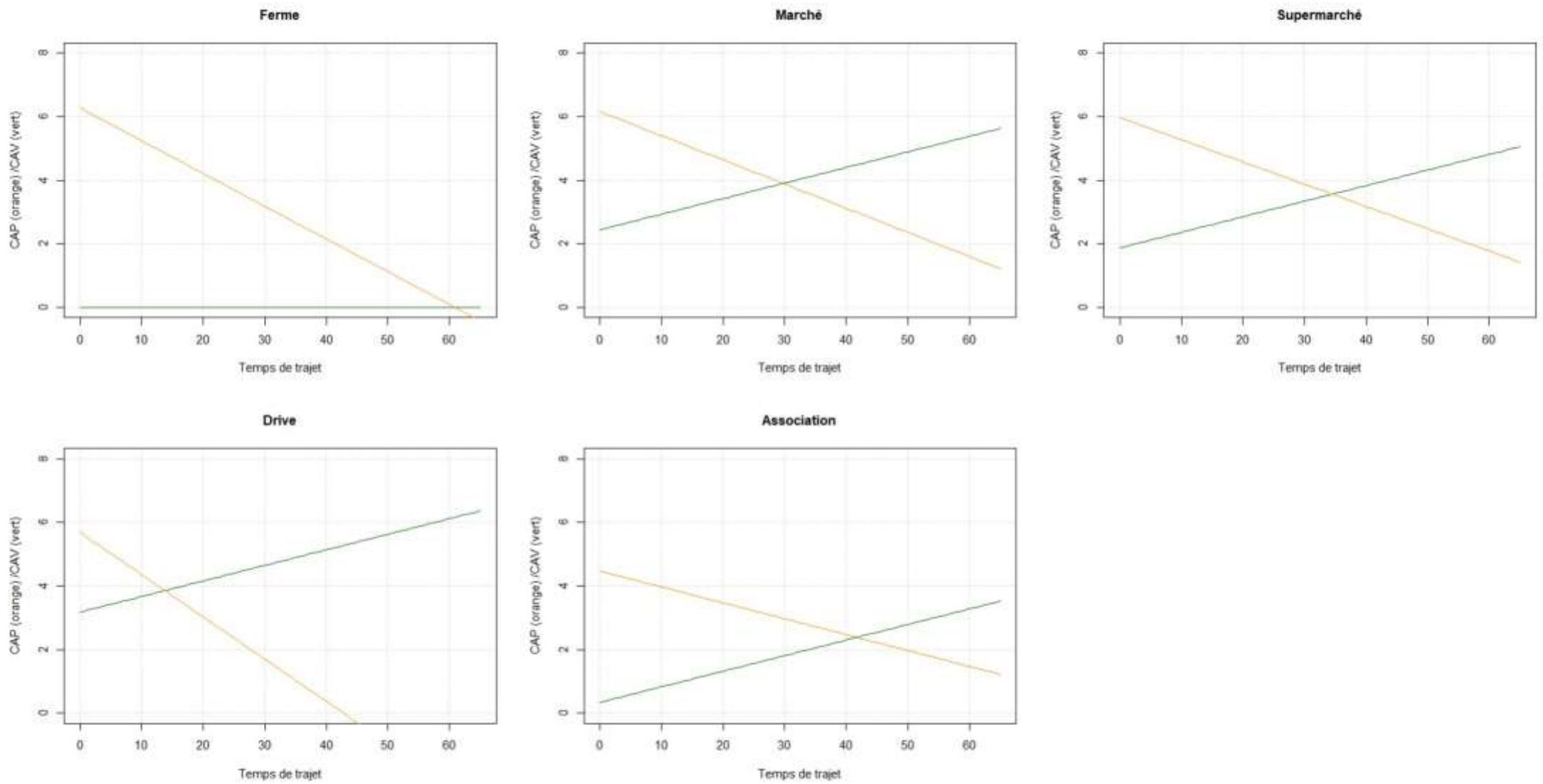

*Figure 1 : Ecarts entre CAP généralisé et CAV généralisé en fonction du temps de trajet, selon les points de vente, pour tous les consommateurs.*

### 5.2 – Pour les consommateurs des CC

Si l'on s'intéresse aux profils les plus engagés (consommateurs achetant en CC), nous obtenons les courbes de l'Annexe 7. Nous pouvons voir que les CAPG sont globalement plus élevés, et sont au-dessus du CAVG pour des temps de trajet plus important. Par exemple, l'association, où consommateurs et agriculteurs acceptent de réaliser jusqu'à 50 minutes de trajet, contre 40 pour l'échantillon entier. Pour le supermarché et le drive, on peut aussi voir que les consommateurs et les agriculteurs acceptent des temps de trajet plus longs, et les consommateurs des prix plus élevés.

### 5.3 – Effets des perceptions

Pour les consommateurs qui déclarent préférer ne pas acheter en CC au supermarché, les résultats sont sur la Figure 3. On voit que leur CAPG sont plus élevés que la moyenne, et qu'eux favorisent certains types de points de vente, en particulier l'association de consommateurs et le marché. Le drive a aussi un potentiel de marché plus important. Le potentiel de marché pour le supermarché est aussi plus important (presque 50min, 4€/kg). Le potentiel de marché pour l'association est plus grand que celui de la ferme.

Pour les consommateurs qui déclarent que les CC permettent de soutenir les agriculteurs locaux, les courbes sont visibles en Annexe 8. Les écarts entre CAPG et CAVG sont plus élevés que la moyenne, mais les préférences pour les points de vente restent les mêmes.

### 5.4 – Pour les consommateurs non-acheteurs des circuits courts

Nous pouvons voir sur la Figure 3 que les écarts entre CAPG et CAVG sont réduits pour les non-acheteurs des CC. La vente au drive ne devient possible que s'ils se trouvent à moins de 10 minutes. Un potentiel de marché existe pour l'association de consommateurs, mais pour des prix moins élevés et des temps de trajets réduits (40min, 2€/kg).

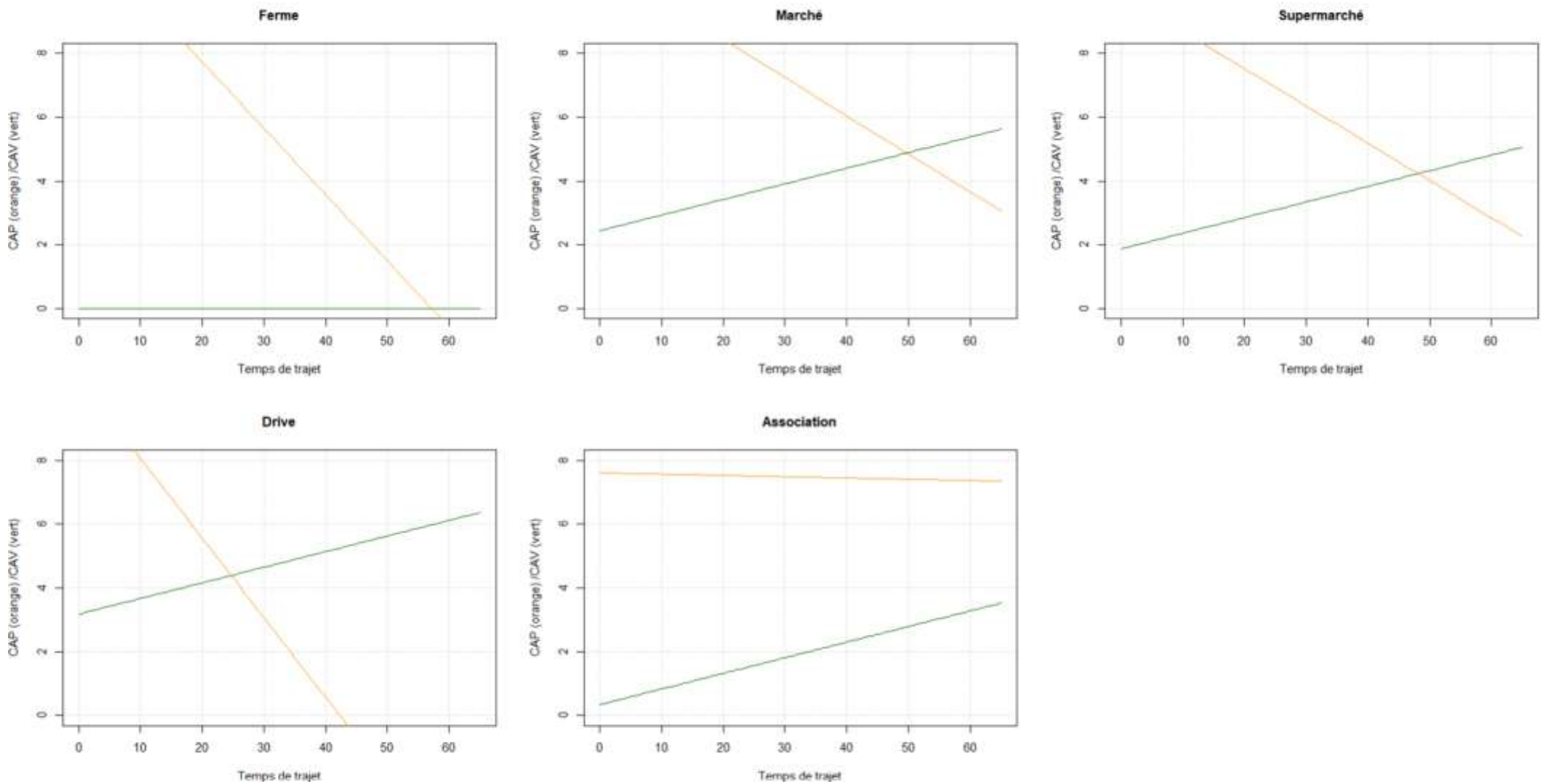

*Figure 2 : CAP généralisé des consommateurs en accord avec la préférence « Je préfère ne pas acheter des produits des CC au supermarché, CAV généralisé*

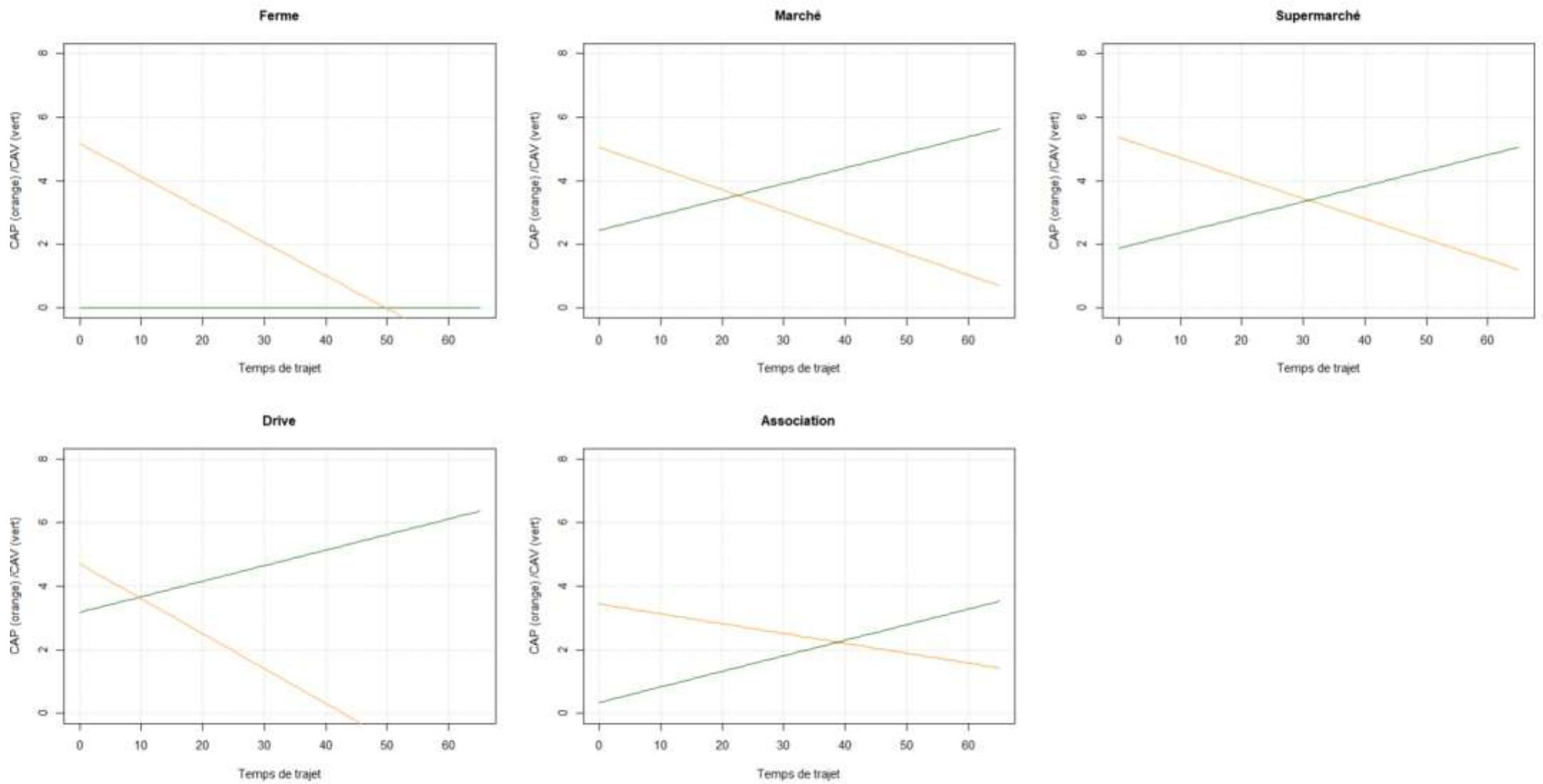

*Figure 3 : CAP généralisé des non-acheteurs des CC, CAV généralisé*

## 6 – Discussion et conclusion

Les travaux actuels pouvant mettre en évidence des caractéristiques d'alternativité des CC se concentraient sur la production, la consommation ou, quand ils abordaient la logistique, sur l'action collective ( Raimbert, Raton, 2021) et la performance socio-territoriale (Raimbert, Raton, 2023). Ces résultats ouvrent des perspectives intéressantes sur le rôle et les conditions de l'engagement des agriculteurs et des consommateurs à participer à la logistique des CC, illustrant de l'intérêt de qualifier ces fonctions logistiques d'alternatives.

L'alternativité proposée par les CC passe par la mise en avant de différentes formes de proximité. Pour commencer, notre travail montre un intérêt des consommateurs sondés pour les produits des CC, y compris par les consommateurs non-acheteurs des CC. Cet intérêt peut être lié à une forme de proximité identitaire, à un engagement moral (Amilien et al. 2022), lié au choix d'acheter en CC plutôt qu'en circuits longs. L'engagement physique (Amilien et al., 2022) peut être mis en avant lorsqu'on s'intéresse aux préférences des consommateurs. Ceux qui ont déclaré être d'accord avec le fait qu'acheter en CC permet de soutenir les agriculteurs locaux ont eu des CAP globalement plus élevés que la moyenne, sans autre changement. En revanche, ceux qui ont traduit cet engagement moral par un acte physique (la non-fréquentation de supermarché), ont non seulement des CAP plus élevés, mais favorisent aussi d'autres points de vente (l'association, la ferme, le marché). Cela peut montrer une certaine polarité des consommateurs, entre les plus engagés, qui privilégient l'association de consommateurs, et les non-acheteurs des CC, qui privilégient le supermarché. Si l'on s'intéresse aux maraîchers, on voit qu'ils ont globalement favorisé les points de vente les plus alternatifs (la ferme, l'association), et ont moins apprécié le drive. On voit aussi que ceux qui ne vendent pas en CC ont un CAV très élevé pour la ferme, mais que les autres points de vente sont potentiellement envisageables. On retrouve donc aussi cette polarité, entre les profils les plus « alternatifs » qui privilégient la ferme et l'association de consommateurs, et les « moins alternatifs », qui recherchent à vendre hors de la ferme. Le supermarché apparait alors comme un repoussoir pour les profils « alternatifs », à la fois pour les consommateurs et les agriculteurs, qui vont préférer l'association de consommateurs et la vente à la ferme. En revanche, pour les autres profils, et en particulier pour les non-acheteurs des CC et les agriculteurs ne vendant qu'en CL, la vente en CC au supermarché peut être vue comme une porte d'entrée à l'achat en CC.

En plus de la proximité identitaire, les consommateurs les plus engagés valorisent significativement la proximité relationnelle avec les maraîchers. Ce lien se trouve à travers les choix de points de vente (ferme, association), mais aussi avec la variable d'organisation d'évènements à la ferme, pour laquelle le CAP est positif et significatif pour les consommateurs sondés. Les maraîchers quant à eux valorisent cette proximité avec les consommateurs en favorisant les points de vente qui leur permettent d'échanger avec eux, mais la variable d'organisation n'est pas significative, ou avec un effet contre-intuitif pour les agriculteurs vendant à la fois en CC et CL. Cela peut être dû au fait que les agriculteurs préfèrent que le temps d'échange soit lié à une activité de vente, et ne représente pas un temps supplémentaire. En revanche, les agriculteurs ont valorisé une forme de proximité relationnelle entre agriculteurs, à travers l'entraide.

Une autre forme de proximité permise par les CC est la proximité d'accès. En évaluant celle-ci à travers les CAP et les CAV pour les temps de trajet, nous pouvons voir que cette proximité d'accès est surtout importante pour le drive et pour l'association. Néanmoins, pour les profils les plus « alternatifs », cette forme de proximité semble s'effacer en partie face à la proximité identitaire, les consommateurs les plus engagés acceptant des temps de trajets élevés pour les points de vente les plus alternatifs. Pour les maraîchers, la faible prise en compte des temps de trajet ressort lorsqu'on distingue le temps de trajet selon le point de vente.

Pour approfondir notre analyse, il serait pertinent d'examiner les décisions des agriculteurs qui n'optent pas pour la commercialisation en CC. Cela permettrait d'obtenir un échantillon plus représentatif de l'ensemble des agriculteurs. De plus, une autre limite de notre évaluation du potentiel de marché réside dans le fait que nous n'avons pas pris en considération l'offre déjà présente sur le territoire où se situent les agriculteurs. Ainsi, pour mieux appréhender ce potentiel de marché, il serait intéressant d'étudier les dynamiques d'entraide et de concurrence qui opèrent dans ces territoires.

**Bibliographie**

**Annexes**

Annexe 1 – Description des attributs de l'enquête (Horvath, 2023) portant sur les proximités relationnelles et fonctionnelles pour les consommateurs puis pour les agriculteurs

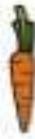 Il n'y a que des légumes vendus dans le point de vente.

Il y a des légumes ainsi que d'autres produits alimentaires (viande, produits laitiers, miel, etc.) vendus dans le point de vente.

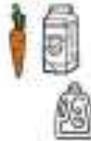 Il y a des légumes, des produits alimentaires (viande, produits laitiers, miel, etc.) et des produits non-alimentaires (produits ménagers etc.) vendus dans le point de vente.

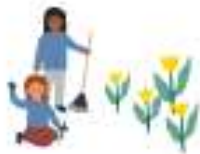 Les agriculteurs de ce point de vente proposent régulièrement aux intéressés des ateliers dégustation, des journées à la ferme pour rencontrer les agriculteurs, et des aides bénévoles sur le terrain pour les agriculteurs (cueillettes, aide pour les cultures…).

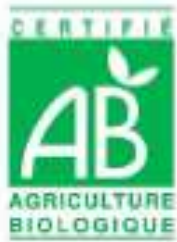 Les légumes sont issus de l'agriculture biologique.

**Signification des logos utilisés dans l'étude :**

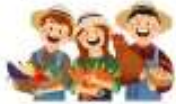

Il y a d'autres agriculteurs autour de vous qui utilisent ce débouché et sont prêts à vous aider (conseils, partage de matériel si besoin, etc)

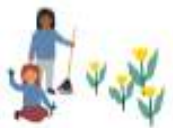

Des consommateurs de ce point de vente aimeraient organiser des journées d'échanges (pique-nique, dégustations, etc) avec les autres consommateurs et les agriculteurs, afin d'en savoir plus sur vos pratiques, de pouvoir travailler sur la ferme bénévolement, de pouvoir visiter la ferme, etc etc…

Annexe 2 – Statistiques descriptives des sous-groupes de consommateurs (1) et de maraîchers (2)

(1)

|  | Tous | | Acheteurs des CC (A) | | Non-acheteurs des CC (B) | | Préf.1 (C) | | Préf.2 (D) | |
|---|---|---|---|---|---|---|---|---|---|---|
|  | Moyenne | Ecart-type | Moyenne | Ecart-type | Moyenne | Ecart-type | Moyenne | Ecart-type | Moyenne | Ecart-type |
| Femmes (%) | 66 |  | 64 |  | 69 |  | 57 |  | 66 |  |
| Age (années) | 43 | 15 | 42 | 14 | 44 | 15 | 38 | 14 | 44 | 15 |
| Diplôme supérieur hors agriculture (%) | 47 |  | 57 |  | 53 |  | 56 |  | 56 |  |
| Actifs (%) | 67,45 |  | 75 |  | 60 |  | 74 |  | 67 |  |
| Revenus du foyer (€/personne/mois) | 2642 | 1096 | 2700 | 1500 | 2600 | 1000 | 2500 | 2100 | 2500 | 2000 |
| Retraités (%) | 14 |  | 11 |  | 14 |  | 9 |  | 13 |  |

(2)

| Maraîchers | Tous | | CC (1) | | CL (2) | | CC/CL (3) | | Préf.CC (4) | |
|---|---|---|---|---|---|---|---|---|---|---|
|  | Moyenne | Ecart-type | Moyenne | Ecart-type | Moyenne | Ecart-type | Moyenne | Ecart-type | Moyenne | Ecart-type |
| Femmes (%) | 36 |  | 44 |  | 36 |  | 36 |  | 44 |  |
| Age (années) | 41 | 11 | 41 | 10 | 41 | 11 | 41 | 11 | 41 | 10 |
| Diplôme supérieur hors agriculture (%) | 43 |  | 69 |  | 58 |  | 43 |  | 69 |  |
| Durée installation (années) | 11 | 11 | 9 | 10 | 10,5 | 10,5 | 11 | 11 | 9 | 10 |
| SAU maraîchage (ha) | 6,58 | 10,94 | 3 | 5 | 10 | 15 | 6,58 | 10,94 | 3 | 5 |
| SAU totale (ha) | 41,56 | 67,3 | 5 | 8 | 15 | 15 | 41,56 | 67,3 | 5 | 8 |

Annexe 3 – Utilités des consommateurs (MNL), réf = non-choix

|  | Tous | CC (A) | CL (B) | Préf. 1 (C) | Préf. 2 (D) |
|---|---|---|---|---|---|
| Ferme | 4.219*** | 4.618*** | 4.266*** | 3.702*** | 4.820*** |
|  | (0.142) | (0.193) | (0.227) | (0.345) | (0.167) |
| Marché | 3.904*** | 4.302*** | 3.968*** | 3.139*** | 4.477*** |
|  | (0.182) | (0.241) | (0.299) | (0.425) | (0.213) |
| Supermarché | 3.953*** | 4.168*** | 4.373*** | 2.999*** | 4.501*** |
|  | (0.189) | (0.248) | (0.311) | (0.435) | (0.219) |
| Drive | 3.992*** | 4.422*** | 4.013*** | 3.438*** | 4.637*** |
|  | (0.170) | (0.223) | (0.284) | (0.387) | (0.200) |
| Association | 2.957*** | 3.462*** | 2.795*** | 2.309*** | 3.510*** |
|  | (0.192) | (0.252) | (0.315) | (0.443) | (0.227) |
| Prix | -0.175*** | -0.154*** | -0.213*** | -0.081*** | -0.192*** |
|  | (0.007) | (0.008) | (0.011) | (0.015) | (0.008) |
| TT Ferme | -0.073*** | -0.064*** | -0.088*** | -0.068*** | -0.074*** |
|  | (0.006) | (0.008) | (0.011) | (0.014) | (0.007) |
| TT_Marché | -0.053*** | -0.050*** | -0.057*** | -0.039*** | -0.059*** |
|  | (0.007) | (0.009) | (0.011) | (0.016) | (0.008) |
| TT Supermarché | -0.049*** | -0.049*** | -0.054*** | -0.038*** | -0.053*** |
|  | (0.006) | (0.008) | (0.009) | (0.015) | (0.007) |
| TT Drive | -0.093*** | -0.093*** | -0.094*** | -0.081*** | -0.103*** |
|  | (0.009) | (0.011) | (0.015) | (0.018) | (0.011) |
| TT Association | -0.035*** | -0.039*** | -0.027. | -0.001 | -0.047*** |
|  | (0.010) | (0.013) | (0.017) | (0.021) | (0.012) |
| Organisation d'évènements | 0.181*** | 0.205*** | 0.141*** | 0.168*** | 0.192*** |
|  | (0.030) | (0.038) | (0.049) | (0.067) | (0.034) |
| Bio | 0.421*** | 0.445*** | 0.375*** | 0.421*** | 0.450*** |
|  | (0.034) | (0.043) | (0.057) | (0.074) | (0.040) |
| Gamme | 0.239*** | 0.250*** | 0.198*** | 0.209*** | 0.292*** |
|  | (0.045) | (0.056) | (0.076) | (0.096) | (0.052) |
| Adjusted R² | 0,10 | 0,12 | 0,11 | 0,10 | 0,13 |
| AIC | 19723 | 11489 | 7951 | 3643 | 14529 |
| BIC | 19817 | 11576 | 8033 | 3713 | 14619 |
| Nombre obs. | 6126 | 3630 | 2496 | 1116 | 4662 |

. p < 0.1, * p < 0.05, ** p < 0.01, *** p < 0.001

Consommateurs – Préf.1 = "Je préfère ne pas acheter en CC au supermarché", Préf.2 = "Acheter en CC permet de soutenir les agriculteurs locaux"

Annexe 4 – Utilités des maraîchers (MNL)

|  | Tous | CC (1) | CL (2) | CC/CL (3) | Préf.CC (4) |
|---|---|---|---|---|---|
| Ferme | 0.404* | 0.499 | -1.841. | 0,65. | 0.601. |
|  | (0.231) | (0.394) | (0.801) | (0,48) | (0.281) |
| Marché | -0.402 | -0.661 | -0.940 | -0,13 | -0.620 |
|  | (0.356) | (0.666) | (1.111) | (0,58) | (0.445) |
| Supermarché | -0.413 | -0.295 | -1.101 | -0,24 | -0.592 |
|  | (0.329) | (0.573) | (1.013) | (0,60) | (0.424) |
| Drive | -0.722* | -1.428* | -1.040 | -0,39 | -1,051* |
|  | (0.344) | (0.658) | (0.976) | (0,56) | (0.445) |
| Association | 0.073 | 0.034 | -0.485 | 0,27 | 0,228 |
|  | (0.344) | (0.578) | (1.121) | (0,56) | (0.423) |
| Prix | 0.189*** | 0.139* | 0.263* | 0,21** | 0.176*** |
|  | (0.048) | (0.083) | (0.162) | (0,068) | (0.059) |
| TT Marché | -0.012 | -0.014 | -0.030 | -0,0069 | -0.0031 |
|  | (0.012) | (0.024) | (0.039) | (0,017) | (0.015) |
| TT Supermarché | -0.006 | -0.020 | -0.003 | -0,00059 | -0.008 |
|  | (0.011) | (0.020) | (0.031) | (0,011) | (0.015) |
| TT Drive | -0.003 | 0.017 | -0.011 | -0,0053 | 0.0082 |
|  | (0.013) | (0.026) | (0.033) | (0,020) | (0.017) |
| TT Asso | -0.015. | 0.0001. | -0.058 | -0,022. | -0.020* |
|  | (0.011) | (0.017) | (0.041) | (0,015) | (0.013) |
| Organisation d'évènements | -0.087 | 0.127 | -0.305 | -0,26* | -0.122 |
|  | (0.078) | (0.138) | (0.263) | (0,10) | (0.097) |
| Entraide | 0.203*** | 0.174 | 0.055 | 0,29** | 0.321*** |
|  | (0.074) | (0.127) | (0.260) | (0,095) | (0.093) |
| Adjusted R² | 0,07 | 0,11 | 0,05 | 0,08 | 0,11 |
| AIC | 3107 | 1062 | 330 | 1565 | 2038 |
| BIC | 3165 | 1108 | 360 | 1615 | 2091 |

|  | Tous | CC (1) | CL (2) | CC/CL (3) | Préf.CC (4) |
|---|---|---|---|---|---|
| Nombre obs. | 924 | 324 | 90 | 468 | 630 |

Annexe 5 – MNL – Utilité des maraîchers sans distinction de temps de trajet (TT) par point de vente

|  | Tous | CC (1) | CL (2) | CC/CL (3) | Préf.CC (4) |
|---|---|---|---|---|---|
| Ferme | -0,38 | 0,47 | -1,87* | 0,62. | 0,53. |
|  | (0,31) | (0,50) | (1,11) | (0,47) | (0,38) |
| Marché | -0,48. | -0,81* | -1,12 | -0,12 | -0,60. |
|  | (0,33) | (0,60) | (1,18) | (0,48) | (0,42) |
| Supermarché | -0,37 | -0,63 | -0,76 | -0,10 | -0,66. |
|  | (0,35) | (0,52) | (1,14) | (0,56) | (0,43) |
| Drive | -0,62* | -1,03* | -0,85 | -0,34 | -0,79* |
|  | (0,33) | (0,50) | (1,08) | (0,51) | (0,41) |
| Association | -0,066 | 0,090 | -1,17. | -0,0077 | -0,074 |
|  | (0,33) | (0,54) | (0,91) | (0,50) | (0,41) |
| Prix | 0,20*** | 0,15* | 0,28. | 0,22*** | 0,19** |
|  | (0,050) | (0,09) | (0,18) | (0,068) | (0,064) |
| TT | -0,0096* | -0,0045 | -0,023. | -0,0093. | -0,0078. |
|  | (0,0047) | (0,0083) | (0,016) | (0,0066) | (0,0057) |
| Organisation d'évènements | -0,070 | 0,11 | -0,23 | -0,19* | -0,086 |
|  | (0,066) | (0,11) | (0,19) | (0,094) | (0,084) |
| Entraide | -0,19** | 0,15 | 0,0014 | 0,28** | 0,31*** |
|  | (0,074) | (0,14) | (0,27) | (0,097) | (0,094) |

Annexe 6 – MNL – Utilité des maraîchers avec distinction de temps de trajet (TT) par point de vente

|  | Tous | CC | CL | CC/CL | Préf.CC |
|---|---|---|---|---|---|
| Ferme | -2,13 | -3,58 | 7,00* | -3,11 | -3,42 |

|  | Tous | CC | CL | CC/CL | Préf.CC |
|---|---|---|---|---|---|
|  | (2,09) | (5,66) | (3,21) | (2,95) | (3,16) |
| Marché | 2,12 | 4,75 | 3,57 | 0,63 | 3,53 |
|  | (1,87) | (4,29) | (3,18) | (2,70) | (2,57) |
| Supermarché | 2,18 | 2,12 | 3,96 | 1,15 | 3,37 |
|  | (1,82) | (4,27) | (4,05) | (2,67) | (2,42) |
| Drive | 3,82** | 10,25* | 4,19. | 1,84 | 5,98** |
|  | (1,96) | (7,38) | (2,82) | (2,50) | (2,8) |
| Association | -0,39 | -0,25 | 1,84 | -1,33 | -1,3 |
|  | (2,12) | (5,03) | (4,55) | (2,88) | (3,05) |
| TT Marché | 0,062* | 0,10 | 0,11 | 0,033 | 0,02 |
|  | (0,06) | (0,14) | (0,16) | (0,080) | (0,09) |
| TT Supermarché | 0,030 | 0,14 | 0,01 | 0,0028 | 0,05 |
|  | (0,05) | (0,17) | (0,08) | (0,052) | (0,08) |
| TT Drive | 0,020 | -0,12 | 0,04 | 0,025 | -0,05 |
|  | (0,070) | (0,23) | (0,13) | (0,091) | (0,5) |
| TT Asso | 0,08. | 0,0001* | 0,22 | 0,11. | 0,11. |
|  | (0,06) | (0,0001) | (0,27) | (0,074) | (0,08) |
| Organisation d'évènements | 0,46 | -1,25 | 1,16 | 1,07** | 0,69 |
|  | (0,43) | (1,27) | (1,46) | (0,61) | (0,6) |
| Entraide | -1,07*** | -0,91 | -0,21 | -1,38** | -1,83** |
|  | (0,48) | (0,99) | (1,05) | (0,66) | (0,87) |
| Nombre obs. | 924 | 324 | 90 | 468 | 630 |

Annexe 7 : Ecarts entre CAP et CAV généralisés pour les sondés acheteurs des C

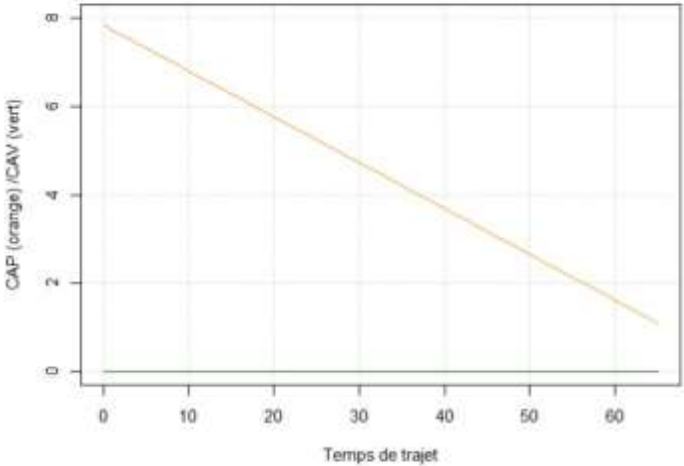
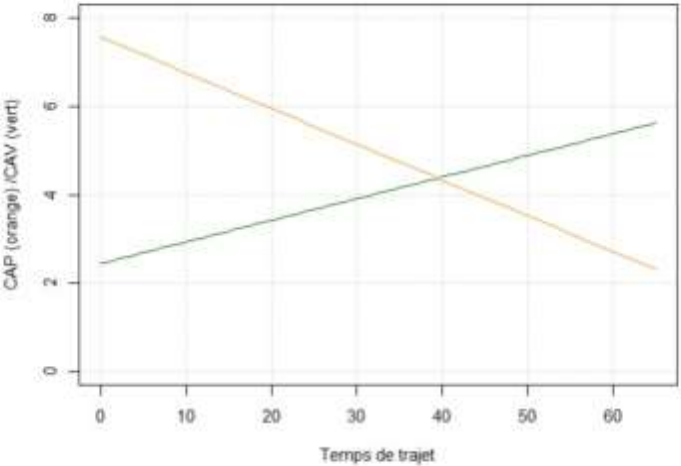
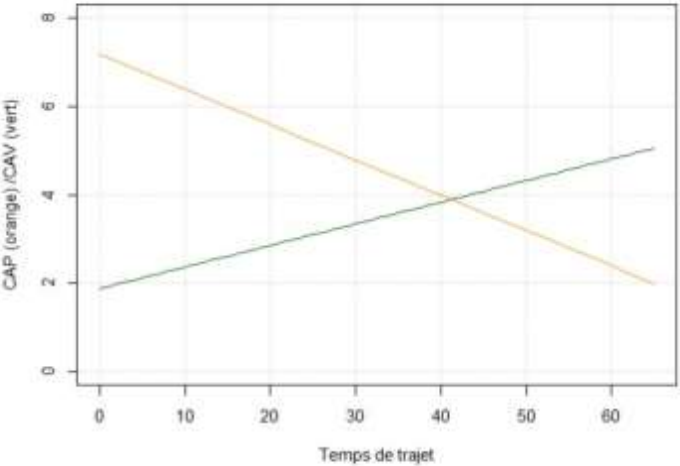
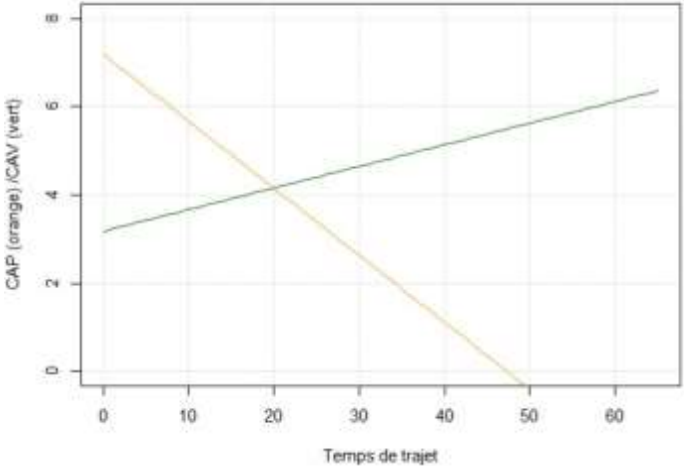
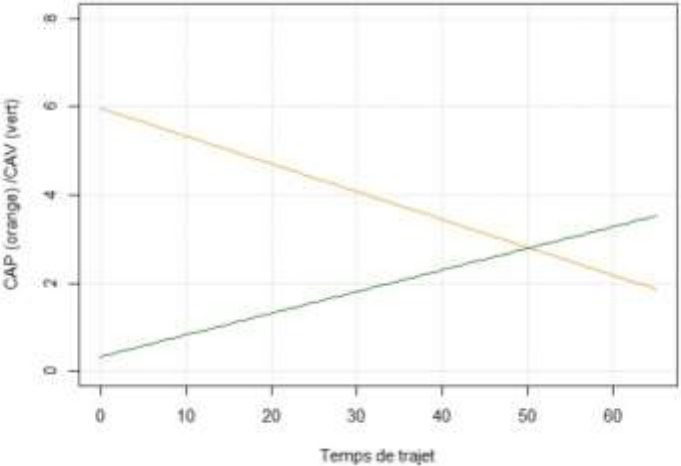

Annexe 8 : Ecarts entre CAV et CAP généralisés pour les sondés qui ont déclaré être en accord avec la préférence « Les CC permettent de soutenir l'agriculture locale »

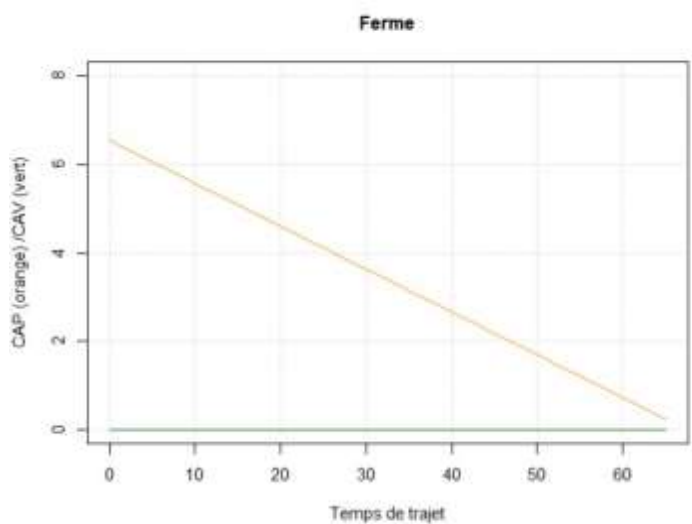
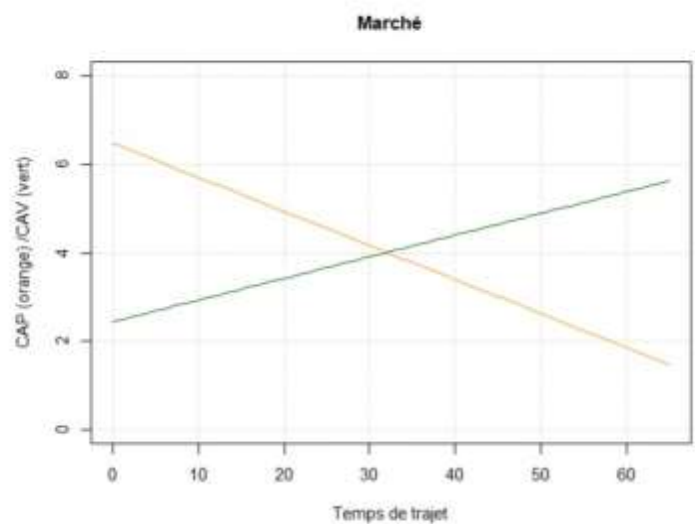
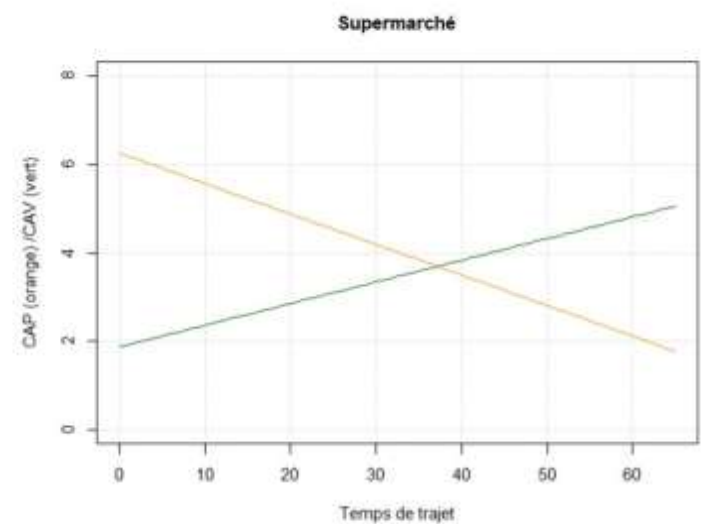
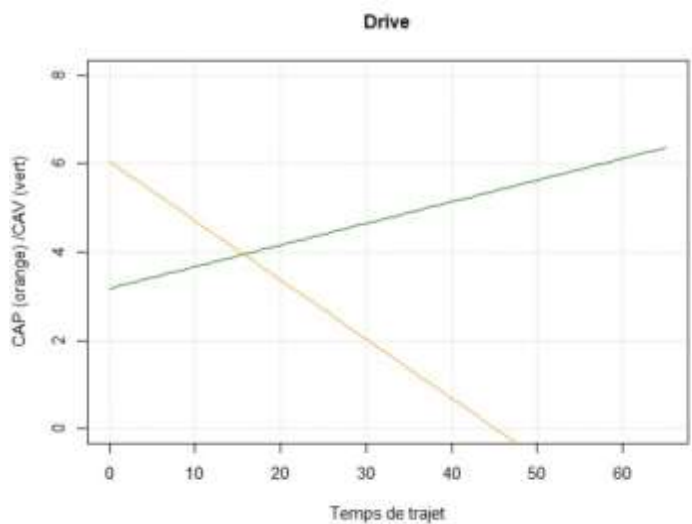
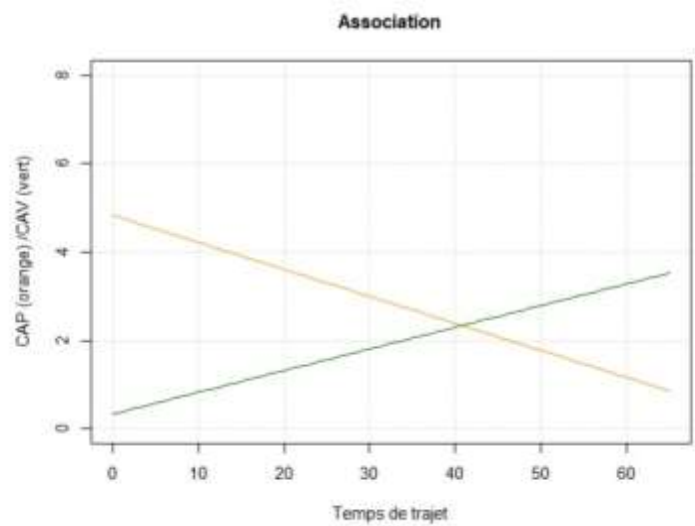